\documentclass[aps,epsf,preprint]{revtex4}

\def\1{{\bf 1}}
\def\[{\left[}
\def\]{\right]}
\def\be{\begin{eqnarray}}
\def\ee{\end{eqnarray}}
\def\nn{\nonumber}
\def\({\left(}
\def\){\right)}

\def\eq#1{(\ref{#1})}

\def\o{\omega}

\def\G{{\cal G}}
\def\l{\lambda}

\def\Tr{{\rm Tr}}
\def\r2{\sqrt{2}}
\def\n{\noindent}
\def\C{{\cal C}}

\begin{document}

\title{The Unique Horizontal Symmetry of Leptons}
\author{C.S. Lam}
\address{Department of Physics, McGill University\\
 Montreal, Q.C., Canada H3A 2T8\\
and\\
Department of Physics and Astronomy, University of British Columbia,  Vancouver, BC, Canada V6T 1Z1 \\
Email: Lam@physics.mcgill.ca}

\begin{abstract}
There is a group-theoretical connection between fermion mixing matrices and minimal horizontal symmetry groups. Applying this connection to the tri-bimaximal neutrino
mixing matrix, we show that the minimal horizontal symmetry group for leptons is uniquely $S_4$, the permutation group of four objects.
\end{abstract}
\narrowtext
\maketitle

\section{Introduction}
Much expectation  is currently being focused on the discovery of the Standard-Model Higgs boson, and on possible experimental indications of what lies beyond. While we await these exciting results, we should not lose sight of the generation problem of more than 70 years, of why there are three generations of fermions and what are the relations between them, for which a certain amount of data is already available to guide us. If (horizontal) symmetry is to play a role in the generation problem just as it does in the Standard Model,
in supersymmetry, in grand unified and string theories, then the generation problem may even have a direct impact on the Higgs boson being hunted for. This is so because, like other symmetries, the horizontal symmetry is expected to be spontaneously broken to yield the  measured quark and lepton masses and mixings. Additional Higgs bosons introduced to break the horizontal symmetry contribute to the mass matrices  and hence the fermion masses, making the fermion coupling of whichever Higgs bosons that will be  first discovered no longer proportional to its mass. That means the estimated fusion production cross section of the Standard-Model Higgs will be off, and its expected decay branching ratio to different fermion pairs will not be as predicted. If that is indeed observed, it would be a strong support for the idea of a horizontal  symmetry.

Many horizontal symmetry groups have been proposed in the literature, among them the groups $Z_m$\cite{Z}, $Z_m\times
Z_n$\cite{ZZ},\ $D_n$\cite{D,LAM},\ $S_3$\cite{S3},\ $S_4$\cite{S4},\ $A_4$\cite{A4},\ $T'$\cite{T},\  $\Delta(27)$\cite{D27}, $SO(3)$ and $SU(3)$\cite{SOU3}. These models introduce a number of additional Higgs bosons to break the spontaneous symmetry, each of which has an adjustable vacuum expectation value and adjustable Yukawa coupling constants. It is  the tuning of these parameters
that allow the same piece of experimental data to be explained by so many different models based on many different symmetries.

This is somewhat unsatisfactory because if Nature possesses  a horizontal symmetry, it must be unique, but how should we determine which of these is the correct symmetry? In a previous letter \cite{CSL},  I suggested what seems to me to be a natural criterion: the correct symmetry group should reveal itself
experimentally without the adjustment of any dynamical parameter. It was further argued that neutrino mixing should be the piece of data used to obtain such a symmetry, whereas
quark mixing and fermion masses have to be obtained from a model-dependent way from a dynamical model based on this symmetry. With this criterion, it was shown that the tri-bimaximal
neutrino mixing matrix \cite{HPS} leads to the symmetry group $S_4$. It was further claimed that this group is unique, in the sense that any other viable symmetry group must 
contain it as a subgroup. The purpose of this paper is to show this uniqueness, as well as other details not fully discussed in \cite{CSL}. Dynamical models based on $S_4$ to
implement quark mixing and fermion masses are not unique and their discussion will not be included in this article.

Since we want to stay away from the adjustment of parameters, the symmetry must be obtained from group theoretical arguments, not dynamical models. The group theoretical
technique needed for this purpose was developed in \cite{LAM}, used there and in \cite{CSL} to obtain the $S_4$ horizontal symmetry
from the tri-bimaximal mixing matrix. For completeness this result will be reviewed in Sec.~II. The proof of the uniqueness of $S_4$ will be carried
out in Sec.~III, by eliminating all other finite subgroups of $SO(3)$ and $SU(3)$ one by one. Sec.~IV is devoted to a brief summary and conclusion.

\section{symmetry and mixing}
We wish to derive the horizontal symmetry from mixing, and vice versa. Since the mixing matrix depends only on left-handed fermions, 
instead of the $3\times 3$ charged-lepton mass matrix $M_e$, which connects left-handed to right-handed charged leptons, it is simpler to deal with $\bar M_e:= M_eM_e^\dagger=\bar M_e^\dagger$, which connects left-handed charged leptons on both sides. Assuming the active
neutrinos to be Majorana, their mass matrix $M_\nu=M_\nu^T$ does not involve right-handed neutrinos and can  be used as it is. 

If $U_e$ and $U_\nu$ are unitary matrices making $U_e^\dagger \bar M_eU_e:= \bar m_e$ and $U_\nu^TM_\nu U_\nu:= m_\nu$ diagonal, then $U=U_e^\dagger U_\nu$ is the neutrino mixing matrix. Since lepton masses are all different, these unitary matrices are essentially unique up to phases. To see that, suppose
$d_e$ and $d_\nu$ are unitary matrices such that $U_ed_e$ also diagonalizes $\bar M_e$ and $ U_\nu d_\nu $ also diagonalizes $M_\nu$,
then $d_e^\dagger \bar m_e d_e=\bar m_e$ and $d_\nu^Tm_\nu d_\nu=m_\nu$, so $d_e$ and $d_\nu$ must both be diagonal, with $d_\nu^Td_\nu=d_\nu^2=1$. In other words, $U_e$
is unique up to an overall phase change of each of its three columns, and $U_\nu$ is unique up to an overall sign change of each of its three columns.

If $F$ is a unitary transformation of the left-handed charged leptons, then this transformation is a horizontal symmetry of the charged leptons iff $F^\dagger \bar M_eF=\bar M_e$. Similarly, if $G$
is a unitary transformation of the active neutrinos, then it is a horizontal symmetry of the active neutrinos iff $G^T M_\nu G=M_\nu$. These equalities imply that $FU_e$  also diagonalizes $\bar M_e$ and
$GU_\nu$ also diagonalizes $M_\nu$, consequently they satisfy $FU_e=U_ed_e$ and $GU_\nu=U_\nu d_\nu$.
The condition $d_\nu^2=1$ implies $G^2=1$. Without loss of generality, we shall confine ourselves to the situation when $\det(G)=1$, so that
$G$ has one $+1$  and two $-1$ eigenvalues. In other words, the columns of $U_e$ are eigenvectors of $F$, and the columns of $U_\nu$ are eigenvectors of $G$ with eigenvalue $+1$ or $-1$. The diagonalization
matrices $U_e$ and $U_\nu$, and hence the mixing matrix $U$, are thereby intimately related to the symmetry operations $F$ and $G$. It is for this reason that the mixing matrix
can determine the symmetry, and vice versa. 

If the Hamiltonian has an unbroken symmetry, this symmetry must be simultaneously a horizontal symmetry for  the charged leptons and the active neutrinos, thus $F=G$. This implies
$U_e=U_\nu$ up to inconsequential phases, and hence $U=1$. In order to have a mixing, $F=G$ is not allowed, so whatever horizontal symmetry that
is present at high energies must be broken at the present energy down to the residual symmetries $F$ for the charged leptons and $G$ for the active neutrinos. With $F\not= G$, the minimal horizontal symmetry group $\G$
at high energy is the (finite) group generated by $F$ and $G$. We shall denote this by the notation $\G=\{F,G\}$.

From now on we choose a basis in which $\bar M_e$ is diagonal. That means $U_e=1$ and  $U=U_\nu$. $F$ is diagonal in that basis because it commutes with $\bar M_e$ whose eigenvalues are all different. Conversely, if $F$ is diagonal {\it and} non-degenerate, in 
the sense that all its eigenvalues are different, then $\bar M_e$ must be diagonal as well. This property will be needed to recover the mixing matrix $U$ from
$F$ and $G$, hence we shall assume $F$ to be non-degenerate from now on .

Given that $U=U_\nu$, $GU=Ud_\nu$,  it is easy to construct $G$ from the neutrino mixing matrix $U$. In fact, there are three solutions $G_i\ (i=1,2,3)$, 
such that $G_iU=Ud_\nu^{(i)}$, with $+1$ occupying the $(ii)$ position of $d_\nu^{(i)}$ and $-1$ occupying the other two diagonal entries. If the mixing matrix is chosen to
have the tri-bimaximal form \cite{HPS},  
\be
U={1\over\sqrt{6}}\pmatrix{2&\sqrt{2}&0\cr -1&\sqrt{2}&\sqrt{3}\cr -1&\sqrt{2}&-\sqrt{3}},\label{tribi}\ee
a form which is well within one standard deviation of the experimental mixing angles, then
\be
G_1&=&{1\over 3}\pmatrix{1&-2&-2\cr -2&-2&1\cr -2&1&-2\cr},\quad
G_2=-{1\over 3}\pmatrix{1&-2&-2\cr -2&1&-2\cr -2&-2&1\cr},\quad
G_3=-\pmatrix{1&0&0\cr 0&0&1\cr 0&1&0\cr}.\label{G}\ee
There is no need to consider $G_1$ from now on because $G_1=G_2G_3$.

In summary, the minimal horizontal group given by the neutrino mixing matrix $U$
is the finite group $\G$ generated by $F,\ G_2$, and $G_3$:\ $\G=\{F,G_2,G_3\}$. The matrices $G_2$ and $G_3$ are given in \eq{G}; the matrix $F$ is unitary and diagonal but otherwise arbitrary.

Conversely, given a horizontal group $\G$ for leptons, whether it can yield the tri-bimaximal mixing \eq{tribi} without fine-tuning depends on whether we can choose three elements $F,\ G_2$, and $G_3$ in $\G$ to act as the residual symmetries after breaking, so that $F$ is non-degenerate, and $G_2$, $G_3$ are given by \eq{G}
in the basis where $F$ is diagonal. If this can be done, then the mixing is given by \eq{tribi} because 
$G_iU=Ud_\nu^{(i)}$ and $\bar M_e$ is diagonal in that basis as remarked earlier.
If no such triplets $F,G_2,G_3$ can be found, then we cannot obtain \eq{tribi} from $\G$ without fine-tuning the dynamical parameters.

Since $\G$ is assumed to be a finite group, there must be a natural integer $n$ such that $F^n=1$. We require $n\ge 3$ in order to keep $F$ non-degenerate. For $n=3$, each entry of the diagonal $F$ must be a different cube
root of unity. Let $\o=e^{2\pi i/3}$, then there are six possible $F$'s, given by $F={\rm diag}(1,\o,\o^2):= F_1$ and the other five permutations of the three entries.
Correspondingly there should be six groups $\G=\{F,G_2,G_3\}$. However, since each $G_i$ is invariant under a simultaneous permutation of its second and third columns
and rows, a permutation of the (22) and (33) entries of $F$ will not give rise to a new group. In this way we cut down the six possible $\G$ to three, the other two being generated by $F_2={\rm diag}(\o,1,\o^2)$ and $F_3={\rm diag}(\o^2,1,\o)$. However, since $F_2^2=F_3$ and $F_3^2=F_2$, these two must generate the same group, so altogether there are only two distinct groups $\G$ from the six possible $F$'s. These groups can be explicitly calculated
by repeatedly multiplying the three generators. The result is:  $\G=\{F_1,G_2,G_3\}=S_4,\ \G=\{F_2,G_2,G_3\}=3.S_4$, where $S_4$ is the symmetric (permutation)
group of four objects with 24 elements, which is also the symmetry group of the cube and of the octahedron, and $3.S_4$ is a 72-element group consisting of
$S_4,\ \o S_4$, and $\o^2 S_4$. Since $3.S_4\supset S_4$, the horizontal symmetry group for $n=3$ is uniquely $S_4$, unique in the sense that any other viable horizontal
group must contain it as a subgroup.

It remains to show the uniqueness of $S_4$ for $n>3$. For a given $n$, there are $n!/(n-3)!$ possible $F$'s, and there is an infinite number of $n$'s, thus
a straight-forward calculation by direct multiplication for every case is clearly impossible. Instead, we will provide a proof  in the next section in another way, 
by enumerating and rejecting the other relevant finite groups.

\section{uniqueness of $S_4$}
In this section we shall show that either $\G=\{F,G_2,G_3\}$ contains $S_4$ as a subgroup, or else it must have an infinite order.

To do that, we first assume $\det(F)=1$. Since $\det(G_2)=\det(G_3)=1$, and since $F,G_2,G_3$ are $3\times 3$ unitary matrices, $\G$ is clearly a 
finite subgroup of $SU(3)$ or $SO(3)$, or, a finite subgroup of $SU(3)$ or $SU(2)$ since $SU(2)$ covers $SO(3)$ twice. The desired conclusion
is reached by enumerating and considering all the
finite subgroups of $SU(2)$ and $SU(3)$, which are known. At the end of the section, we will show that the conclusion remains valid when the unit-determinant assumption of $F$ is dropped.

\subsection{Finite Subgroups of $SU(2)$ and $SU(3)$}
The finite subgroups of $SU(2)$ are in one-one correspondence with the simply-laced Lie algebras $A_n, D_n, E_6, E_7, E_8$ \cite{MBD, B, MK}. The two infinite series
$A_n$ and $D_n$ correspond respectively to the cyclic group $Z_n$ with $n$ elements, and the dihedral group $D_n$ with $2n$ elements. The three exceptional groups
$E_6, E_7, E_8$ correspond to the three symmetry groups ${\cal T}, {\cal O}, {\cal I}$ of the regular polyhedrons, with ${\cal T}$ the symmetry of the tetrahedron,
${\cal O}$ the symmetry of the cube and the octahedron, and ${\cal I}$ the symmetry of the dodecahedron and the icosahedron. They are also equal to the 
groups $A_4, S_4, A_5$ respectively, where $S_n$ is the symmetric group, namely, the permutation group of $n$ objects, and $A_n$ is the alternating group, namely, all the even permutations of $n$ objects.

The finite subgroups of $SU(3)$ not in $SU(2)$ are also known \cite{MBD,B,FFK,BLW,LNR,LNR2}; the notations used below are those of \cite{FFK}, in which the number within the parentheses is the order of the group. There are also two infinite series, $\Delta(3n^2)$ and $\Delta(6n^2)$, and six `exceptional' ones: $\Sigma(36), \Sigma(60), \Sigma(72), \Sigma(168), \Sigma(216), \Sigma(360)$.

We shall show in the subsequent paragraphs that these groups cannot yield the tri-bimaximal mixing matrix \eq{tribi} without parameter tuning unless they contain $S_4$ as a subgroup.
We shall do that by dividing these groups into different categories.

\subsection{Groups without three-dimensional irreducible representations}
It was shown in \cite{LAM} that the horizontal symmetry group has to possess a three-dimensional (3D) irreducible 
representation (IR), or else we can never recover the tri-bimaximal mixing without a tuning of parameters.
This is so because of the particular form of \eq{G}, and because the left-handed fermions must belong to
a 3DIR to avoid the presence of tunable parameters.

The series of groups $Z_n$, being Abelian, has only 1DIR, and the series of groups $D_n$
has at most 2DIR, so both are unsuitable. The groups $\Sigma(36), \Sigma(72)$, and $\Sigma(360)$ do not have 3DIR 
either \cite{FFK}, so they must be rejected as well. 

\subsection{$A_4$}
This subgroup of $S_4$ gives automatically trimaximal mixing, but it can give bimaximal mixing as well only
by tuning some Yukawa coupling constants \cite{MA}. To obtain tri-bimaximal mixing without any tuning, we need the
rest of the symmetries contained in $S_4$.

\subsection{$\Delta(3m^2)$ and $\Delta(6m^2)$}
These two infinite series are discussed in \cite{MBD, B, FFK, BLW, LNR, EL}, with their 3DIR's given
in \cite{BLW, LNR, EL}. We must show that none of them are allowed unless they contain $S_4$ as 
a subgroup.

There are many 3DIR's, but every one of those matrices has only one non-zero element in each row and each column, and  these non-zero elements
all have absolute value 1. This is the only property that is required to rule out these two series.

There are $3!=6$ types of matrices with this property. They are
\be
t1&:=&\pmatrix{\rho_1&0&0\cr 0&0&\sigma_1\cr 0&\tau_1&0},
t2:=\pmatrix{0&0&\rho_2\cr 0&\sigma_2&0\cr \tau_2&0&0},
t3:=\pmatrix{0&\rho_3&0\cr \sigma_3&0&0\cr 0&0&\tau_3},\nn\\
t4&:=&\pmatrix{\rho_4&0&0\cr 0&\sigma_4&0\cr 0&0&\tau_4},
t5:=\pmatrix{0&\rho_5&0\cr 0&0&\sigma_5\cr \tau_5&0&0},
t6:=\pmatrix{0&0&\rho_6\cr \sigma_6&0&0\cr 0&\tau_6&0},\label{type}\ee
with $|\rho_i|=|\sigma_i|=|\tau_i|=1$. The question is whether we can pick from them three
members $g_2,g_3,f$ which can be turned into $G_2,G_3$ and $F$ by a unitary 
transformation $V$, such that $F$ is diagonal, non-degenerate, and having an order $n>3$.  $n=1,2$ are excluded
because we need the three eigenvalues of $F$ to be different; $n=3$ is not considered because $G_2,G_3,F$ simply
generate $S_4$. Unless three such members can be picked, there will be no new horizontal symmetry contained in these two series of groups.

$G_2$ and $G_3$ can be simultaneously diagonalized by the tri-bimaximal mixing matrix $U$ of \eq{tribi} so that 
\be d_\nu^{(2)}=U^\dagger G_2 U=\pmatrix{-1&0&0\cr 0&1&0\cr 0&0&-1\cr},\quad d_\nu^{(3)}=U^\dagger G_3 U=\pmatrix{-1&0&0\cr 0&-1&0\cr 0&0&1\cr}\label{G23d} \ee
are diagonal. To find $V$, we first determine the unitary matrix $v$ which simultaneously diagonalizes $g_2$ and $g_3$ 
into diagonal forms
$g_{2d}=v^\dagger g_2v$ and $g_{3d}=v^\dagger g_3 v$, then identify $g_{2d}$ with $d_\nu^{(2)}$ and $g_{3d}$
with $d_\nu^{(3)}$. The desired matrix $V$ so that $V g_{2,3}V^\dagger=G_{2,3}$ is then given by $V=Uv^\dagger$.

Such a matrix $v$ exists because $g_2$ and $g_3$ commute, for otherwise $G_2$ and $G_3$ would not have commuted. 
Let $v_a\ (1\le a\le 6)$ be the unitary matrix which renders $t_{ad}=v_a^\dagger t_av_a$ diagonal, then $v$ must be one
of these $v_a$'s.

Since $\Tr(G_2)=\Tr(G_3)=-1$,  we must require $\Tr(g_2)=\Tr(g_3)=-1$. Since $n>3$ and the three eigenvalues
of $F$ are required to be different, we also have $\Tr(F)\not=0$, and hence $\Tr(f)\not=0$.
The trace of any matrix in $t_5$ and $t_6$ is zero, thus we know that $g_2,g_3,f$ must all come from $t_i$ for $1\le i\le 4$.
To have the right trace, we need to have $\rho_1=\sigma_2=\tau_3=\rho_4+\sigma_4+\tau_4=-1$. To have unit determinant which $G_2$ and $G_3$ possess,
we also require $\sigma_1\tau_1=\rho_2\tau_2=\rho_3\sigma_3=-1$.

Incidentally, this analysis excludes $\Delta(24)=S_4$ which has $n=3$. In that case $\Tr(f)=0$ and its $f$ belongs to
type $t_5$ or $t_6$.

The unitary diagonalization matrices $v_a$ of the first four types can be taken to be
\be v_1={1\over\r2}\pmatrix{\r2&0&0\cr 0&w_1&-w_1\cr 0&1&1\cr}, v_2={1\over\r2}\pmatrix{w_2&0&-w_2\cr
0&\r2&0\cr 1&0&1}, v_3={1\over\r2}\pmatrix{w_3&-w_3&0\cr 1&1&0\cr 0&0&\r2\cr},\ee
with $|w_i|=1$. $v_4$ can be taken to be the identity matrix.

It is easy to check from \eq{type} that the vanishing of the commutators between $g_2$ and $g_3$
requires either (i) both of them to belong to the same type $t_i$, or (ii) one of them belongs to type $t_4$. 
For $j=1,2,3$, the $(jj)$ matrix element of $t_j$ and $t_{jd}$ is $-1$, so we see from \eq{G23d} that when (i)
occurs, both $g_2$ and $g_3$ must belong to type $t_1$ or type $t_4$. Let us consider the different cases separately.

\vspace{.5cm}
\n \underline{(ia). both $g_2$ and $g_3$ belong to type $t_1$}

In order for $t_{1d}=v_1^\dagger t_1v_1$ to be diagonal,
$w_1\tau_1=w_1^*\sigma_1$ is required, in which case
\be
t_{1d}=\pmatrix{\rho_1&0&0\cr 0&\tau_1w_1&0\cr 0&0&-\tau_1w_1\cr}.\ee
This can be identified with $d_\nu^{(2)}$ and $d_\nu^{(3)}$ in \eq{G23d} by setting $\rho_1=-1$ and $w_1=\pm 1/\tau_1$. 

Now that we know $V=Uv^\dagger$, we can compute $VfV^\dagger$ to see whether it can be identified with $F$,
which is a diagonal non-degenerate matrix with order $n>3$. Since $f$ has to be in one of the four types
$t_1,t_2,t_3,t_4$, we can calculate them all. The result is that this can never happen no matter what parameters we choose.

\vspace{.5cm}
\n \underline{(ib). both $g_2$ and $g_3$ belong to type $t_4$}

In that case $v$ is the identity and $V=U$. Again we can verified that none of $Ut_iU^\dagger$ for $i\le 4$ can be the 
desired $F$.

\vspace{.5cm}
\n \underline{(ii). one of $g_2$ and $g_3$ belongs to $t_4$ and the other to $t_i$}

We shall deal with the case when $g_2$ is of type $t_4$. The other case when $g_3$ is in $t_4$ is very similar.

Comparing with \eq{G23d}, we see that $\rho_4=-1=-\sigma_4=\tau_4$. If $g_3$ belongs to type $t_i$, its diagonalization
matrix $v_i$ must diagonalize $g_{2}$ as well, hence $i=2$. As in case (ia) above, $V=Uv^\dagger=Uv_2^\dagger$
can then be computed and $VfV^\dagger$ checked whether it can lead to a desired $F$. The result is the same before,
no matter which one of the four types $t_i$ that $f$ belongs to, it can never gives rise to the desired $F$. 

This concludes the proof that $\Delta(3m^2)$ and $\Delta(6m^2)$ cannot give rise to a new horizontal symmetry
not containing $S_4$.

\subsection{$A_5$, $\Sigma(168)$, $\Sigma(216)$}
The remaining three groups, the icosahedral group $A_5=\Sigma(60)$, the Klein group $\Sigma(168)=L_2(7)$ \cite{LNR2},
and the Hessian group $\Sigma(216)$, can all be ruled out by using their class structures and character tables.
To explain how this is done let us denote any of these three groups by $\G$.

The question is whether an equivalent representation can be chosen
to enable three members $f,g_2,g_3\in\G$ to be picked out  to be equal to $F,G_2.G_3$,
respectively. If so, $\G$ is capable of giving the tri-bimaximal mixing. If not,  $\G$ can be ruled out.
Such an identification requires $f$ to have a finite order $n\ge 3$, and $g_2, g_3$ to have order 2,
and be given by the formula \eq{G}.

Since the proof relies only on characters and eigenvalues, which are the same for all equivalent representations,
we do not have to worry about which equivalent representation we choose.

We need not consider those $f$ with $n=3$ because they would never give us anything new.  
If we can find $g_{2,3}$ that are equal to $G_{2,3}$, then $\G$ must contain $S_4$
as a subgroup, so $S_4$
remains to be the only minimal horizontal group compatible with tri-bimaximal mixing without tuning.
If not, then $\G$ can never give rise to tri-bimaximal mixing so it is ruled out.

For $n>3$, we rely on the following strategy to rule out these groups $\G$. It is shown below how class
structure and character table can be used to determine the eigenvalues of $f=F$. This tells us what the diagonal
forms of $F$ are. If $g_{2,3}$ can
be found to equal to $G_{2,3}$ in this representation of diagonal $F$, then  
the group generated by $\{F,G_2,G_3\}$ must be a subgroup of $\G$. 
However, we shall show that in each case when $n>3$, the element $FG_2$ and/or the element
$FG_3$ has an order larger than the order of the group $\G$, hence this group cannot be a subgroup
these groups $\G$,  there can be no $g_{2,3}$ that can be found to be identified with $G_{2,3}$. This is how
these groups $\G$ are ruled out.

\subsubsection{$A_5$}
The character table from \cite{FFK} is given in Table I, where $b_{5+}=(-1+\sqrt{5})/2$ 
and $b_{5-}=(-1-\sqrt{5})/2$. The first row of the table names the 5 conjugacy
classes, the second row gives the permutation structure of each class in cycle notations of $S_5$. 
For example, $1^23$ consists of
two 1-cycles and one 3-cycle. The third row tells us the order of the elements, for example, class $\C_4$
consists of elements of order 5 and their fourth powers, and $\C_5$ consists of the second and third powers of 
elements of order 5. $E$ is the identity element. 
The next five rows are the characters of the irreducible representations of each class, with the
boldface numerals in the first column giving the dimension of the representation. The last two rows will be explained
later.
\begin{center}
{\bf Table I: character table for $A_5$} \\
\vspace{.5cm}
\begin{tabular}{c|ccccc}
classes&$\C_1$&$\C_2$&$\C_3$&$\C_4$&$\C_5$\\
perm type& $1^5$&$1^23$&$12^2$&$5$&$5$\\ 
elem type& $E$&($C_3, C_3^2$)&$C_2$&$(C_5, C_5^4)$&$(C_5^2, C_5^3)$ \\ \hline
{\bf 1}& 1&1&1&1&1\\ 
{\bf 3}& 3&0&$-1$&$-b_{5-}$&$-b_{5+}$\\
${\bf 3'}$&3&0&$-1$&$-b_{5+}$&$-b_{5-}$\\
{\bf 4}&4&1&0&$-1$&$-1$\\
{\bf 5}& 5&$-1$&1&0&0\\ \hline
$\C^2$&$\C_1$&$\C_2$&$\C_1$&$\C_5$&$\C_4$\\
$\C^3$&$\C_1$&$\C_1$&$\C_3$&$\C_4$&$\C_5$\\

\end{tabular}
\end{center}

In principle, $F$ can be taken to be any of the 60 elements of the group, but to get anything other than $S_4$, 
we merely have to concentrate on those elements with order $n>3$. This leaves 
elements in class $\C_4$ or $\C_5$, both of order $n=5$. To determine the eigenvalues in each class, 
we list in the last two rows of the
table what class the square and the cube of every $\C_i$ belong to. 
Once this is known, the eigenvalues of every element $F$ can be deduced from $Tr(F), Tr(F^2)$ and $Tr(F^3)$, namely,
from the character table.

Let us illustrate how the last two rows are obtained. For example,
 class $\C_4$ consists of elements of the type $C_5$ and $C_5^4$, hence the square of any element $f\in\C_4$
is of the form $C_5^2$ and $C_5^3$, so $f^2\in\C_5$. Similarly, if $f\in\C_5$, then $f^2\in\C_4$. 
 We will now proceed
to determine the eigenvalues of $F$ from the characters. 

Let $\o_n=\exp(2\pi i/n)$. Then it is well known that
\be \sum_{i=0}^{n-1}\o_n^i=0.\label{phasesum}\ee
Using that for $n=5$, it will now be shown that $b_{5+}=\o_5+\o_5^4$, and $b_{5-}=\o_5^2+\o_5^3$. To start with, 
let $x=\o_5+\o_5^4$, and $y=\o_5^2+\o_5^3$. Then \eq{phasesum} tells us that $x+y=-1$. Now
$x^2=y+2=-x+1$, and $y^2=x+2=-y+1$. Thus both $x$ and $y$ satisfy the quadratic equation $z^2+z=1$, whose 
two solutions are $z=(-1\pm\sqrt{5})/2=b_{5\pm}$, so one must be $x$ and the other must be $y$. To determine which is which,
note that $\o_5$ is in the first quadrant of the complex plane and $\o_5^4$ is in the fourth quadrant, so $x$ must have
a positive real part. Thus $x=b_{5+}$ and
$y=b_{5-}$.

Knowing that, it is now easy to verify that if $F\in\C_4$, then its eigenvalues in the {\bf 3} representation
are $1, \o_5^2, \o_5^3$. These are also
the eigenvalues of $F$ in the ${\bf 3'}$ representation if $F\in \C_5$. Similarly, 
the eigenvalues for any $F\in\C_5$ in the {\bf 3}
representation and any $F\in\C_4$ in the ${\bf 3'}$ representation, are $1, \o_5, \o_5^4$. 

Given three distinct eigenvalues, $3!=6$ diagonal $F$'s can be produced, depending on where the eigenvalues
are put. In order to show that $h=FG_2$ or $h=FG_3$ has an order larger than 60, the order of $A_5$, we compute its three
eigenvalues $\l_1, \l_2, \l_3$ numerically. Let us assume $|\l_i|=1$, for otherwise $h$ cannot have a finite order.
In that case, $\l_i=exp(2\pi i \theta_i)$, and the numerical value for each $\theta_i$ is approximated by a
rational number $k_i/m_i$. Then $h$ would have an order larger than 60 if $m_i>60$, and that turns out
to be true in every case.

\subsubsection{$\Sigma(168)$}
This group is studied in great detail in \cite{LNR2}, but for our present purpose, it is sufficient just to use
the character table taken from \cite{FFK}:

\begin{center}
{\bf Table II: character table for $\Sigma(168)$} \\
\vspace{.5cm}
\begin{tabular}{c|ccccccc}
classes&$\C_1$&&$\C_2$&$\C_3$&$\C_4$&$\C_5$&$\C_6$\\
perm type& $1^7$&&$1^32^2$&$124$&$13^2$&$7$&7\\ 
elem type& $E$&&$C_2$&$(C_4,C_4^3)$&$(C_3, C_3^2)$&$(C_7,C_7^2, C_7^4)$&$(C_7^3,C_7^5,C_7^6)$ \\ \hline
{\bf 1}& 1&&1&1&1&1&1\\ 
{\bf 3}& 3&&$-1$&$1$&0&$b_{7+}$&$b_{7-}$\\
${\bf 3^*}$&3&&$-1$&$1$&0&$b_{7-}$&$b_{7+}$\\
{\bf 6}&6&&2&0&0&$-1$&$-1$\\
{\bf 7}& 7&&$-1$&$-1$&1&0&0\\ 
{\bf 8}&8&&0&0&$-1$&1&1\\                  \hline
$\C^2$&$\C_1$&&$\C_1$&$\C_2$&$\C_4$&$\C_5$&$\C_6$\\
$\C^3$&$\C_1$&&$\C_2$&$\C_3$&$\C_1$&$\C_6$&$\C_5$\\

\end{tabular}
\end{center}

This table is listed in the same way as the table for $A_5$, with $b_{7\pm}=(-1\pm i\sqrt{7})/2$.
From this table we see that candidates for $F$ with $n>3$ should come from classes $\C_3, \C_5$, or $\C_6$.
Therefore we need to figure out the eigenvalues of elements in these classes.

For $\C_3$, since the order of its elements is $n=4$, the three eigenvalues of $F$ have to be chosen from
the four values $\pm 1, \pm i$. Both the {\bf 3} and ${\bf 3^*}$ characters of $\C_3$ are 1, and the $\C_2$
characters are $-1$. Thus if $F\in\C_3$, then $Tr(F)=Tr(F^3)=1$ and $Tr(F^2)=-1$, so the eigenvalues
are $1, +i, -i$.

The character of $\C_5$ is $b_{7+}$ in {\bf 3} and $b_{7-}$ in ${\bf 3}^*$. It is the reverse for $\C_6$.
Let us now prove that $b_{7+}=\o_7+\o_7^2+\o_7^4$ and $b_{7-}=\o_7^3+\o_7^5+\o_7^6=b_{7+}^*$. The proof is
similar to the case of $b_{5\pm}$. Letting $x=\o_7+\o_7^2+\o_7^4$ and $y=\o_7^3+\o_7^5+\o_7^6$, it follows that
both $x$ and $y$ satisfy the quadratic equation $z^2+z+2=0$, whose solutions are $b_{7\pm}$. Moreover, $x$
should be identified with $b_{7+}$ because their imaginary parts are both positive.

With this relation we can now determine the eigenvalues of $F$ in $\C_5$ and $\C_6$, both for {\bf 3} and ${\bf 3^*}$. 
For $\C_5$ and {\bf 3} or $\C_6$ and ${\bf 3^*}$,  $Tr(F)=Tr(F^2)=b_{7+}=\o_7+\o_7^2+\o_7^4$ and 
$Tr(F^3)=b_{7-}=\o_7^3+\o_7^5+\o_7^6$. Hence the eigenvalues of $F$ are $\o_7, \o_7^2, \o_7^4$. Similarly,
for $F\in\C_6$ and {\bf 3} or $F\in\C_5$ and ${\bf 3^*}$, the eigenvalues are $\o_7^3, \o_7^5, \o_7^6$.
For each set of eigenvalues, $3!=6$ diagonal matrices $F$ can be produced.

As in the case of $A_5$, $h=FG_2$ or $h=FG_3$ has an order larger than 168 in every case, thereby ruling out $\Sigma(168)$.

\subsubsection{$\Sigma(216)$}
The character table taken from \cite{FFK} is,

\begin{center}
{\bf Table III: character table for $\Sigma(216)$} \\
\vspace{.5cm}
\begin{tabular}{c|ccccccccccc}
classes&$\C_1$&&$\C_2$&$\C_3$&$\C_4$&$\C_5$&$\C_6$&$\C_7$&$\C_8$&$\C_9$&$\C_{10}$\\
perm type& $1^9$&&$1^33^2$&$1^33^2$&$14^2$&$126$&$126$&$12^4$&$3^3$&$3^3$&$3^3$\\ 
elem type& $E$&&$C_3$&$C_3^2$&$(C_4,C_4^3)$&$C_6$&$C_6^5$&$C_2$&$(C_3',C_3^{'2})$&$C_3''$&$C_3^{''2}$ \\ \hline
{\bf 1}& 1&&1&1&1&1&1&1&1&1&1\\ 
${\bf 1'}$&1&&$\o$&$\o^2$&$1$&$\o$&$\o^2$&1&1&$\o$&$\o^2$\\
${\bf 1^{'*}}$&1&&$\o^2$&$\o$&$1$&$\o^2$&$\o$&1&1&$\o^2$&$\o$\\
{\bf 2}&2&&$-1$&$-1$&0&1&1&$-2$&2&$-1$&$-1$\\
${\bf 2'}$&2&&$-\o$&$-\o^2$&0&$\o$&$\o^2$&$-2$&2&$-\o$&$-\o^2$\\
${\bf 2^{'*}}$&2&&$-\o^2$&$-\o$&0&$\o^2$&$\o$&$-2$&2&$-\o^2$&$-\o$\\
{\bf 3}& 3&&0&0&$-1$&$0$&0&3&3&0&0\\
${\bf 8}$&8&&$2$&$2$&0&0&0&0&$-1$&$-1$&$-1$\\
${\bf 8'}$& 8&&$2\o$&$2\o^2$&0&0&0&0&$-1$&$-\o$&$-\o^2$\\
${\bf 8^{'*}}$&8&&$2\o^2$&$2\o$&0&0&0&0&$-1$&$-\o^2$&$-\o$\\                 \hline
$\C^2$&$\C_1$&&$\C_3$&$\C_2$&$\C_7$&$\C_2$&$\C_3$&$\C_1$&$\C_8$&$\C_{10}$&$\C_{9}$\\
$\C^3$&$\C_1$&&$\C_1$&$\C_1$&$\C_4$&$\C_7$&$\C_7$&$\C_7$&$\C_1$&$\C_1$&$\C_1$\\

\end{tabular}
\end{center}
where $\o=\o_3$. To have $n>3$, $F$ must come from $\C_4$, $\C_5$, or $\C_6$. From the {\bf 3}-representation 
row of the character table, we find that
if $F\in\C_4$, then $Tr(F)=Tr(F^3)=-1$ and $Tr(F^2)=3$. The eigenvalues of $F$ are then $1, -1, -1$. Since
two of the three eigenvalues are identical, we must reject this case. If $F\in \C_5$ or $\C_6$, then $Tr(F)=Tr(F^2)=0$ and 
$Tr(F^3)=3$. The eigenvalues are then $1,\o,\o^2$. In spite of having order $n=6$ for elements in $\C_5$, their 3DIR
is identical to an $F$ with $n=3$, whose answer is already known. Hence $\Sigma(216)$ cannot produce anything new.

\subsection{General $F$}
We will now relax the condition $\det(F)=1$. Since $F^n=1$ for some $n$, $\eta:=\det(F)$ is an $n$th root of unity, and
the $3\times 3$ unitary matrix $F$ can be written 
$F=\eta^{1/3} F'$, where  
$\det(F')=1$.

If $\G=\{F,G_2,G_3\}$ forms a finite group, then its presentation \cite{CM, WIKI} is defined by a number of relations $R_a(F,G_2,G_3)=1$, where $R_a$ are monomials of $F, G_2$, and $G_3$.
Three of these relations are $R_1=F^n, R_2=G_2^2$, and $R_3=G_3^2$, but there must be others relating $F, G_2$ and $G_3$. If $F$ appears $k$ times in $R_a$, then
by taking the determinant of the relation on both sides, we see that $\eta^{k}=1$. 

Every group element $g_i$ of $\G$ can be written as a monomial of $F, G_2, G_3$: $g_i=g_i(F,G_2.G_3)$. This monomial is not unique because we can always insert a number of $R_a$ in it. Nevertheless,
if $g_1g_2=g_3$, the sum of powers of $F$ appearing in $g_1$ and $g_2$ must equal to the power of $F$ appearing in $g_3$, modulo $m$, where $m$ is the smallest integer such that $\eta^m=1$.  This conclusion can  be reached by taking the determinant on both sides of the equation.

Now consider the group $\G'=\{F', G_2, G_3\}$ generated by $F', G_2$ and $G_3$. Since $\eta^k=1$, $R_a(F', G_2, G_3)=1$, hence a relation of $\G$ is also a relation of
$\G'$. Moreover, the mapping $g_i\to g_i':=g_i(F',G_2,G_3)$ is a homomorphism from $\G$ to $\G'$, preserving multiplication relations. The kernel of the mapping is a
subgroup ${\cal Z}$ of $\G$, consisting of all elements $z(F,G_2,G_3)$ for which $z':=z(F',G_2,G_3)=1$. If the power of $F$ in the monomial $z$ is $d$, then it follows
that $z=\eta^d 1$. In other words, ${\cal Z}=\{\eta^d 1\}$ is isomorphic to a subgroup of the cyclic group $Z_n$.

Thus $\G$ is a central extension of $\G'$, consisting of elements of the form $zg'$, with $z\in{\cal Z}$ and $g'\in\G'$. Moreover, $\G'$ is a finite subgroup of $SU(3)$.
Since the only group $\G'$ that can naturally lead to the tri-bimaximal mixing is a group containing $S_4$, the same is true for $\G=\{F,G_2,G_3\}$,
so the uniqueness of $S_4$ is established even if 
$\det(F)\not=1$.

\section{conclusion}
We have shown that the horizontal group for leptons is uniquely $S_4$, or any group containing it. To reach this conclusion, we have used the criterion that a horizontal
group should be obtained from the neutrino mixing matrix and vice versa without parameter tuning. To implement this criterion, a purely group-theoretical link between
neutrino mixing matrices and horizontal symmetry groups is established. When this link is applied to the tri-bimaximal neutrino mixing matrix, $S_4$ emerges as a
possible horizontal group. Other finite groups not containing $S_4$ are all ruled out by studying the finite subgroups of $SU(2)$ and $SU(3)$.

Quark mixing and fermion masses are obtained from $S_4$-invariant dynamical models. They are model dependent and their discussion will be postponed to a future publication.

I am grateful to James Bjorken, Ernest Ma, and John McKay for discussions.


\begin{thebibliography}{9}
\bibitem{Z}  Y. Koide, H. Nishiura, K. Matsuda, T. Kikuchi, and T. Fukuyama, Phys.~Rev. D66 (2002) 093006; E. Ma, arXiv:hep-ph/0312192; W. Grimus, A.S.Joshipura, S. Kaneko, L. Lavoura, H. Sawanaka, and M. Tanimoto, Nucl.~Phys. B713 (2005) 151; B. Hu, F. Wu, Y.-L. Wu, Phys.~Rev. D75 (2007) 113003.  

\bibitem{ZZ} N. Haba, C. Hattori, M. Matsuda, and T. Matsuoka, arXiv:hep-ph/9511312; A. Ghosal and D. Majumdar, Phys.~Rev. D66 (2002) 053004; T. Asaka and Y. Takanishi,
arXiv:hep-ph/0409147; S.-L. Chen, M. Frigerio, and E. Ma, Phys.~Lett. B612 (2005) 29; 
W. Krolikowski, Acta Phys.~Polon. B36 (2005) 865; arXiv:hep-ph/0501008; Y. Kajiyama, M. Raidal, and A. Strumia, Phys.~Rev. D76 (2007) 117301;
C. Luhn, S. Nasri, and P. Ramond, Phys.~Letts. B652 (2007) 27.

\bibitem{D} C.D. Carone, R.F. Lebed, Phys.~Rev. D60 (1999) 096002; R. Dermisek and S. Raby, Phys.~Rev. D62 (2000) 015007; Phys.~Lett. B622 (2005) 327; H. Ishimori, T. Kobayashi, H. Ohki, Y. Omura, R. Takahashi, and M. Tanimoto, arXiv:0803.0796; W. Grimus, A.S. Joshipura, S. Kaneko, L. Lavoura, and M. Tanimoto, JHEP 0407 (2004) 078; E. Ma, Fizika B14 (2005) 35;
C. Hagedorn and W. Rodejohann, JHEP 0507 (2005) 034; S.-L. Chen and Ernest Ma, Phys.~Lett. B620 (2005) 151; J. Kubo, Phys.~Lett. B622 (2005) 303; M. Honda, R. Takahashi, and M. Tanimoto, JHEP 0601 (2006) 042; C. Hagedorn, M. Lindner and F. Plentinger, Phys.~Rev. D74 (2006) 025007;  
R. Dermisek, M. Harada, and S. Raby, Phys.~Rev. D74 (2006) 035011; P. Ko, T. Kobayashi, J.-h. Park, and S. Raby, Phys.~Rev. D76 (2007) 035005; D76 (2007) 059901(E);
A. Blum, R. N. Mohapatra, and W. Rodejohann, arXiv:0706.3801; A. Blum, C. Hagedorn and M. Lindner, Phys.~Rev. D77 (2008) 076004;
A. Blum, C. Hagedorn and A. Hohenegger, High Energy Phys. 03 (2008) 070; 
H. Ishimori, T. Kobayashi, H. Ohki, Y. Omura, R. Takahashi, and M. Tanimoto,Phys.~Lett. B662 (2008) 178; Phys.~Rev. D77 (2008) 115005;
H. Okada, arXiv:0804.0926. 



\bibitem{LAM} C.S. Lam, arXiv:0708.3665, Phys.~Lett. B656 (2007) 193.

\bibitem{S3} S. Pakvasa and H. Sugawara, Phys. Lett. B73 (1978) 61;
Y.-P. Yao, arXiv:hep-ph/9507207; L.J. Hall, H. Murayama, Phys.~Rev.~Lett. 75 (1995) 3985; 
K. Kang, J.E. Kim, and P. Ko, Z.~Phys. C72 (1996) 671;
C.D. Carone, Nucl.~Phys.~Proc.~Suppl. 52A (1997) 177;
S.L. Adler, Phys.~Rev. D59 (1998) 015012; Erratum-ibid. D59 (1999) 099902; 
M. Tanimoto, Phys.~Rev. D59 (1999) 017304; Acta Phys.~Polon. B30 (1999) 3105; Phys.~Lett. B483 (2000) 417; 
Y. Koide, Phys.~Rev. D60 (1999) 077301;
E. Ma, Phys.~Rev. D61 (2000) 033012; R.N. Mohapatra, A. Pérez-Lorenzana, and C.A. de S. Pires, Phys.~Lett. B474 (2000) 355; J.I. Silva-Marcos, JHEP 0307 (2003) 012;
P.F. Harrison and W. G. Scott, Phys.~Lett. B557 (2003) 76;
T. Kobayashi, J. Kubo, and H. Terao, Phys.~Lett. B568 (2003) 83; K. Hamaguchi, M. Kakizaki, and M. Yamaguchi, Phys.~Rev. D68 (2003) 056007; J. Kubo, A. Mondragon, M. Mondragon, and E. Rodriguez-Jauregui, Prog.~Theor.~Phys. 109 (2003) 795; Erratum-ibid. 114 (2005) 287; T. Kobayashi, J. Kubo, and H. Terao, Phys.~Lett. B568 (2003) 83;
J. Kubo, Phys.~Lett. B578 (2004) 156; Erratum-ibid. B619 (2005) 387; Phys.~Rev. D70 (2004) 036007; S.-L. Chen, M. Frigerio, and E. Ma, Phys.~Rev. D70 (2004) 073008; Erratum-ibid. D70 (2004) 079905; W.-l. Guo, Phys.~Rev. D70 (2004) 053009; T. Araki, J. Kubo, and E.A. Paschos, Eur.~Phys.~J. C45 (2006) 465; W. Grimus and L. Lavoura, JHEP 0508 (2005) 013; JHEP 0601 (2006) 018; J.~Phys. G34 (2007) 1757; T. Teshima, Phys.~Rev. D73 (2006) 045019; Y. Koide, Phys.~Rev. D73 (2006) 057901; Eur.~Phys.~J. C50 (2007) 8009; J.E. Kim and J.-C. Park, JHEP 0605 (2006) 017; N. Haba, A. Watanabe, and K. Yoshioka, Phys.~Rev.~Lett. 97 (2006) 041601; N. Haba and K. Yoshioka, Nucl.~Phys. B739 (2006) 254; 
H. Morisi, arXiv:hep-ph/0604106; M. Picariello, arXiv:hep-ph/0611189;
R.N. Mohapatra, S. Nasri, and H.-B. Yu, Phys.~Lett. B639 (2006) 318; R.N. Mohapatra and H.-B. Yu, Phys.~Lett. B644 (2007) 346; O. Felix, A. Mondragon, M. Mondragon, and E. Peinado, Rev.~Mex.~Fis. S52 (2006)  67;  A. Mondragon, M. Mondragon, and E. Peinado, Phys.~Rev. D76 (2007) 076003; J.~Phys. A41 (2008) 304035; K.S. Babu, S.M. Barr, and I. Gogoladze, 
Phys.~Lett. B661 (2008) 124; C.-Y. Chen and L. Wolfenstein, Phys.~Rev, D77 (2008) 093009; 
M. Mitra and S. Choubey, arXiv:0806.3254.  

\bibitem{S4} S. Pakvasa and H. Sugawara, Phys. Lett. B82 (1979) 105; Y. Yamanaka, H. Sugawara and S. Pakvasa, Phys.~Rev. D25 (1961) 1895;
E. Ma, hep-ph/0508231, Phys.~Lett. B632 (2006) 352; C. Hagedorn, M. Lindner, and R.N. Mohapatra, JHEP 0606 (2006) 042; 
Y. Cai and H.-B. Yu, Phys.~Rev. D74 (2006) 115005;
B. Lampe, arXiv:hep-ph/0610270; F. Caravaglios and S. Morisi, Int.~J~Mod.~Phys. A22 (2007) 2469; 
H. Zhang, Phys.~Lett. B655 (2007) 132; Y. Koide, J.~High Energy Phys. 08 (2007) 086; arXiv:0707.0899; S. Nandi and Z. Tavartkiladze, Phys.~Lett. B661 (2008) 109; M.K. Parida, Phys.~Rev. D78 (2008) 053004.

\bibitem{A4} E. Ma and G. Rajasekaran, Phys.~Rev. D64 (2001) 113012, Mod.~Phys.~Lett. A16 (2001) 2207; E. Ma, Mod.~Phys.~Lett. A17 (2002) 289, 627;
J.~Phys. G29 (2003) 313; Mod.~Phys.~Lett. A17 (2002) 2361; Phys.~Rev. D70 (2004) 031901; New J.~Phys. 6 (2004) 104; Mod.~Phys.~Lett. A20 (2005) 2767;
Mod.~Phys.~Lett. A20 (2005) 2601; Phys.~Lett. B632 (2006) 352; Phys.~Rev. D73 (2006) 057304; Mod.~Phys~.Lett. A21 (2006) 2931; Mod.~Phys.~Lett. A22 (2007) 101;
arXiv:hep-ph/0701016; arXiv:0808.1729;
K.S. Babu, Ts. Enkhbat, and I. Gogoladze, Phys.~Lett. B555 (2003) 238; K.S. Babu, E. Ma, and J.W.F. Valle, Phys.~Lett. B552 (2003) 207; 
K.S.Babu, T. Kobayashi, and J. Kubo, Phys.~Rev. D67 (2003) 075018; M. Hirsch, J.C. Romao, S. Skadhauge, J.W.F. Valle, and A. Villanova del Moral,
Phys.~Rev. D69 (2004) 093006; G. Altarelli and F. Feruglio, Nucl.~Phys. B720 (2005) 64; Nucl.~Phys. B741 (2006) 215; 
S.-L. Chen, M. Frigerio, and E. Ma, Nucl.~Phys. B724 (2005) 423; M. Hirsch, E. Ma, J.W.F. Valle, and A. Villanova del Moral,
Phys.~Rev. D72 (2005) 091301; Erratum-ibid. D72 (2005) 119904; X.-G. He, Y.-Y. Keum, and R.R. Volkas, JHEP 0604 (2006) 039; 
B. Adhikary, B. Brahmachari, A. Ghosal, E. Ma, and M.K. Parida, Phys.~Lett. B638 (2006) 345; E. Ma, H. Sawanaka, and M. Tanimoto, Phys.~Lett. B641 (2006) 301;
X.-G. He and A. Zee, Phys.~Lett. B645 (2007) 427; B. Adhikary and A. Ghosal, Phys.~Rev. D75 (2007) 073020; arXiv:0803.3582; L. Lavoura and H. Kuhbock, Mod.~Phys.~Lett. A22 (2007) 181;
S.F. King, M. Malinsky, Phys.~Lett. B645 (2007) 351; X.-G. He, arXiv:hep-ph/0612080; Y. Koide, arXiv:hep-ph/0701018; S. Morisi, M. Picariello, and E. Torrente-Lujan,
Phys.~Rev. D75 (2007) 075015; M. Hirsch, A.S. Joshipura, S. Kaneko, and J.W.F. Valle, arXiv:hep-ph/0703046; F. Yin, Physics Review D75 (2007) 073010; 
F. Bazzocchi, S. Kaneko, and S. Morisi, JHEP 03(2008) 063; W. Chao, S. Luo, Z.-Z. Xing, and S. Zhou, arXiv:0709.1069; W. Grimus and H. Kuhbock, Phys.~Rev. D77 (2008) 055008;
F. Bazzocchi, S. Morisi, and M. Picariello, arXiv:0710.2928; L. Lavoura and H. Kuhbock, Eur.~Phys.~J. C55 (2008) 303; M. Honda and M. Tanimoto, arXiv:0801.0181; 
B. Brahmachari, S. Choubey, and M. Mitra, arXiv:0801.3554; G. Altarelli, F. Feruglio, and C. Hagedorn, arXiv:0802.0090; T. Fukuyama, arXiv:0804.2107; Y. Lin, arXiv:0804.2867;
C. Csaki, C. Delaunay, C. Grojean, and Y. Grossman, arXiv:0806.0356; P.H. Frampton and S. Matsuzaki, arXiv:0806.4592; F. Feruglio, C. Hagedorn, Y. Lin, and L. Merlo,
arXiv:0807.3160; H. Ishimori, T. Kobayashi, Y. Omura, and M. Tanimoto, arXiv:0807.4625. 

\bibitem{T} A. Aranda, C.D. Carone, R.F. Lebed, Int.~J.~Mod.~Phys. A16S1C (2001) 896; 
F. Feruglio, C. Hagedorn, Y. Lin, and L. Merlo, arXiv:hep-ph/0702194; P.H. Frampton and T.W. Kephart, arXiv:0706.1186; M.-C. Chen and K.T. Mahanthappa, arXiv:0710.2118;
P.H. Frampton and S. Matsuzaki, arXiv:0710.5928; G.-J. Ding, arXiv:0803.2278.


\bibitem{D27} G.C. Branco, J.M. Gerard, and W.Grimus, Phys.~Lett. B136 (1984) 383;
I. de Medeiros Varzielas, S. F. King, and G. G. Ross, arXiv:hep-ph/0607045; E. Ma, Mod.~Phys.~Lett. A21 (2006) 1917; arXiv:0709.0507;
R. Howl and S.F. King, 	Phys.~Rev. B77  (2008) 144527. 

\bibitem{SOU3} S. F. King and G. G. Ross, Phys.~Lett. B574 (2003) 239; T. Appelquist, Y. Bai, and M. Piai, Phys.~Rev. D74 (2006) 076001;
I. de Medeiros Varzielas, S.F. King, and G.G.  Ross, Phys.~Lett. B644 (2007) 153; Y. Kiode, rXiv:0707.0899; Riazuddin, arXiv:0707.0912;
T.-L. Wu, arXiv:0807.3847; W.-Y.P. Hwang, arXiv:0808.2091.

\bibitem{CSL} C.S. Lam, arxiv:0804.2622, to appear in the Physical Review Letters.

\bibitem{HPS} P.F. Harrison, D.H. Perkins, and W.G. Scott, Phys.~Lett. B458,  (1999) 79,
hep-ph/9904297; Phys.~Lett. B530, (2002) 167, hep-ph/0202074.
\bibitem{CM} H.S.M. Coxeter and W.O.J. Moser, `Generators and Relations for Discrete Groups', (Springer-Verlag,
1980).
\bibitem{WIKI} http://en.wikipedia.org/wiki/Presentation\_ of\_ a\_ group.


\bibitem{MBD} G.A. Miller, H.F. Blichfeldt, and L.E. Dickson, `Theory and applications of finite groups', John Wiley and Sons, 1916.
\bibitem{B} H.F. Blichfeldt, `Finite collineation groups', University of Chicago Press, 1917.
\bibitem{MK} J. McKay, Proc.~Symp.~Pure Math. 37 (1980) 183.
\bibitem{FFK} W.M. Fairbairn, T. Fulton, and W.H. Klink, J.~Math.~Phys. 5, 1038 (1964).
\bibitem{BLW} A. Bovier, M. L\"uling, and D. Wyler, J.~Math.~Phys. 22, 1543 (1981).
\bibitem{LNR} C. Luhn, S. Nasri, and P. Ramond, arXiv:hep-th/0701188, J.~Math.~Phys. 48, 073501 (2007).
\bibitem{EL} J.A. Escobar and C. Luhn, arXiv:0809.0639.
\bibitem{LNR2} C. Luhn, S. Nasri, and P. Ramond, arXiv:0709.1447.


\bibitem{MA} E. Ma, hep-ph/0606039, Pramana 67 (2006) 803.




\end{thebibliography}
\end{document}